\documentclass{jfm}
\usepackage{graphicx}
\usepackage{epstopdf, epsfig}
\usepackage{amsmath}
\usepackage{array}

\usepackage[colorlinks, allcolors=blue]{hyperref} 

\newcommand\Ra{\mbox{\textit{Ra}}}
\newcommand\Cn{\mbox{\textit{Cn}}}

\newcommand\We{\mbox{\textit{We}}}
\newcommand\Fr{\mbox{\textit{Fr}}}

\newcommand\Nun{\mbox{\textit{Nu}}}
\newcommand\Prn{\mbox{\textit{Pr}}} 
\newcommand\Pe{\mbox{\textit{Pe}}}

\shorttitle{Turbulent RB convection with bubbles attached to the plate}
\shortauthor{H.-R. Liu, K. L. Chong, R. Yang, R. Verzicco and D. Lohse}

\title{Turbulent Rayleigh-B\'enard convection with bubbles attached to the plate}

\author{Hao-Ran Liu\aff{1},
  Kai Leong Chong\aff{2},
   Rui Yang\aff{1},
  Roberto Verzicco\aff{3,4,1}
 \and Detlef Lohse\aff{1,5},\corresp{\email{d.lohse@utwente.nl}}}

\affiliation{\aff{1}Physics of Fluids Group and Max Planck Center Twente for Complex Fluid Dynamics,\\
MESA+Institute and J. M. Burgers Centre for Fluid Dynamics, University of Twente,\\
P.O. Box 217, 7500AE Enschede, The Netherlands
\aff{2}Shanghai Key Laboratory of Mechanics in Energy Engineering, Shanghai Institute of Applied Mathematics and Mechanics, School of Mechanics and Engineering Science, Shanghai University, Shanghai, 200072, PR China
\aff{3}Dipartimento di Ingegneria Industriale, University of Rome ``Tor Vergata", Via del Politecnico 1, Roma 00133, Italy
\aff{4}Gran Sasso Science Institute - Viale F. Crispi, 7 67100 L’Aquila, Italy
\aff{5}Max Planck Institute for Dynamics and Self-Organization, Am Fassberg 17, 37077 Göttingen, Germany}

\begin{document}

\maketitle

\begin{abstract}
We numerically investigate turbulent Rayleigh-B\'enard convection with gas bubbles attached to the hot plate, mimicking a core feature in electrolysis, catalysis, or boiling. The existence of bubbles on the plate reduces the global heat transfer due to the much lower thermal conductivity of gases as compared to liquids and changes the structure of the boundary layers. The numerical simulations are performed in 3D at Prandtl number $\Prn=4.38$ (water) and Rayleigh number $10^7\le\Ra\le10^8$. For simplicity, we assume the bubbles to be equally-sized and having pinned contact lines. We vary the total gas-covered area fraction $0.18 \le S_0 \le 0.62$, the relative bubble height $0.02\le h/H \le0.05$ (where $H$ is the height of the Rayleigh-B\'enard cell), the bubble number $40 \le n \le 144$, and their spatial distribution. In all cases, asymmetric temperature profiles are observed, which we quantitatively explain based on the heat flux conservation at each horizontal section. We further propose the idea of using an equivalent single-phase setup to mimic the system with attached bubbles. Based on this equivalence, we can calculate the heat transfer. Without introducing any free parameter, the predictions for the Nusselt number, the upper and lower thermal boundary layer thicknesses, and the mean centre temperature well agree with the numerical results. Finally, our predictions also work for the cases with much larger $\Prn$ (e.g. $400$), which indicates that our results can also be applied to predict the mass transfer in water electrolysis with bubbles attached to the electrode surface or in catalysis.

\end{abstract}


\section{Introduction}

In wall-bounded buoyancy-driven turbulence, the boundary conditions play a crucial role in the flow structure and the global transport property of the system. Bubbles attached to the wall affect these boundary conditions. They often occur in various industrial applications. One example is water electrolysis, where bubbles are generated at the electrodes and can significantly reduce the global mass transport of the system by reducing the active electrode area \citep{vogt2005ea,wang2014rser,yang2018pccp,sepahi2021}, leading to the decrease of the electrolyser efficiency. Another example is catalysis, where bubbles are generated by chemical reactions and can block the catalytic surface, thus also reducing the mass transport \citep{somorjai2010book,oehmichen2010cet,xu2018acr}. One example from daily life is heating water. When a pot of water is heated from below, many tiny gas bubbles nucleate at the bottom wall, reducing the heat transfer efficiency of the system due to the much lower thermal conductivity of gas as compared to the liquid. In all of these examples, bubbles attached to the wall influence the boundary layer (BL) and thus affect the global heat or mass transfer of the system. Therefore, it is highly desirable to quantitatively understand how much the global transport properties are changed. This is the motivation for our study.

As model system we choose Rayleigh-B\'enard (RB) convection which is the paradigm of thermally driven turbulence (see the reviews of \cite{ahl09,loh10,chi12,shishkina2021prf}), where a fluid between two parallel plates is heated from below and cooled from above. In previous studies, the effect of various plate properties on the flow structure and the heat transport were examined, e.g., staggered conducting and insulating strips on the plate \citep{wang2017jfm,bakhuis2018jfm}, temporally-modulated temperature \citep{xia2008jfm,yang2020prl}, plate with roughness \citep{zhu2017prl,jiang2018prl,zhu2019jfm}, and plates with different wettabilities \citep{liu2022jfm}. Here we pick RB convection as model system to study the effects of attaching bubbles on the global heat or mass transport in buoyancy-driven turbulence.

We employ direct numerical simulations with an advanced finite difference method combined with the phase field method \citep{liu2021jcp}. The bubbles are put at the lower (hot) plate with pinned contact lines so that they cannot move or detach from the plate. To focus on the effects of the Rayleigh number (dimensionless strength of the thermal driving) and the bubble geometry, we disregard mass exchange between the bubble and the liquid, and keep the bubbles at constant volume. The calculations are performed for various Rayleigh numbers and geometries, expressed through the relative area covered by the bubbles, bubble height, bubble number, and type of bubble distribution. To define the thermal BL thickness in the multiphase system, we extend the traditional temperature profile slope method, based on heat flux conservation. With this, we propose an equivalent single-phase system, for which we apply the Grossmann-Lohse (GL) theory \citep{gro00,gro01,ste13} to predict the heat transfer and the temperature profile for RB convection with bubbles attached to the hot plate. Without introducing any new free parameter, these predictions well agree with our numerical results. They work for both moderate and large $\Prn$, which indicates that they can be applied to predict both the mass transport in water electrolysis with bubbles attached to the electrode surface and to catalytic surfaces on which bubbles have formed.

The organization of this paper is as follows: The numerical method and setup are introduced in Section \ref{meth}. The flow features and heat transfer are shown in Section \ref{sec1}. We define the thermal BL thicknesses of this new two-phase system in Section \ref{sec-bl}, and propose an equivalent single-phase system in Section \ref{sec2} to calculate the heat transfer. 
The paper ends with conclusions and an outlook.

\section{Numerical method and setup}
\label{meth}

The three-dimensional simulations are performed in a cubic domain of dimensions $H^3$. The numerical method \citep{liu2021jcp} combines the phase-field method \citep{jaqcmin1999jcp, ding2007jcp, liu2015jcp} and an advanced finite difference direct numerical simulation solver for the Navier-Stokes equations \citep{ver96, poe15cf}, called \href{https://github.com/PhysicsofFluids/AFiD}{AFiD}. Numerical details, validation cases, and convergence tests were already presented in our previous study \citep{liu2021jcp}.

The phase field method is widely used in simulations of multiphase turbulent flows \citep{soligo2021jfe}, where the liquid-gas interface is represented by contours of the volume fraction $C$ of the liquid. The corresponding volume fraction of gas is $1-C$. The evolution of $C$ is governed by the Cahn-Hilliard equation,
\begin{equation}
\frac {\partial C} {\partial t} + \nabla \cdot ({\bf u} C) = \frac{1}{\Pe}\nabla^2 \psi,
\label{ch}
\end{equation} 
where $\bf u$ is the flow velocity, and $\psi= C^{3} - 1.5 C^{2}+ 0.5 C  -\Cn^{2} \nabla^2 C$ the chemical potential. We set the P\'eclet number $\Pe=0.9/\Cn$ and the Cahn number $\Cn=0.75\Delta x/H$ with $\Delta x$ being the mesh size and $H$ being the height of the RB cell. The parameters $\Pe$ and $\Cn$ are taken according to the sharp-interface approach proposed in \cite{ding2007jcp,yue2010jfm,liu2015jcp}.

The flow is governed by the Navier-Stokes equation, the heat transfer equation, and the incompressibility condition,
\begin{equation}
\tilde{\rho} \left(\frac {\partial {\bf u}} {\partial t} + {\bf u} \cdot \nabla {\bf u}\right)= - \nabla P + \sqrt {\frac{\Prn}{\Ra}} \nabla \cdot [\tilde{\mu} (\nabla {\bf u}+ \nabla {\bf u}^{T})] + {{\bf F}_{st}}+{\bf G},
\label{ns}
\end{equation}

\begin{equation}
\tilde{\rho}\tilde{c_p} \left(\frac{\partial {\theta}}{\partial t} + {\bf u} \cdot \nabla \theta \right) =   \sqrt{\frac{1}{ \Prn \Ra }} \nabla \cdot (\tilde{k} \nabla \theta),
\label{t}
\end{equation}

\begin{equation}
\nabla \cdot {\bf u}= 0,
\label{con}
\end{equation}
where $\theta$ is the dimensionless temperature, ${\bf F}_{st}=6\sqrt{2}\psi \nabla C / (\Cn \We)$ the non-dimensionalized surface force, where $We=\rho_l U^2 H/\sigma$ is the Weber number, with the surface tension $\sigma$. The vector ${\bf G} = \left\{ [ C + \Lambda_\beta \Lambda_\rho (1-C) ] \, \theta - \tilde{\rho} / \Fr \right\} {\bf z}$ represents the dimensionless gravity. All dimensionless material properties (indicated by a tilde, $\tilde{q}$) are defined in a uniform way, $\tilde{q}=C+\Lambda_q(1-C)$, where $\Lambda_q=q_{g}/q_{l}$ is the ratio of the material properties of gas and liquid, marked by the subscripts $g$ and $l$, respectively. The global dimensionless parameters controlling the flow are listed in Table \ref{table-para}. 
The most important response parameter of the system is the heat transfer, which is quantified by the Nusselt number $\Nun=Q/(k_l\Delta/H)$, with $Q$ being the dimensional heat flux.

The values of the control parameters are chosen mainly based on the properties of air and water (see Table \ref{table-para}), though for better numerical efficiency we take the density ratio $\Lambda_\rho=0.01$, about $10$ times larger than in reality. Note that the exact value of this parameter hardly affects our results. Since the bubbles are pinned and the buoyancy and surface tension forces applied on bubbles are always balanced, we take $Fr=1$ and $We=100$, also for numerical convenience. 
The geometrical parameters of the bubbles are the relative covered area $0.18\le S_0\le0.62$, the non-dimensionalized height $0.02\le h/H\le0.05$, their number $40\le n\le144$, and the spatial distribution (uniform, random and half-covered). Note that from $S_0$, $\tilde{h}=h/H$ and $n$, we can also calculate the bubble volume ($=S_0 \tilde{h}/2+n \pi \tilde{h}^3/6$), the bubble contact radius ($=\sqrt{S_0/(n\pi)}$), the bubble contact angle ($=\arccos\{[S_0/(2n\pi \tilde{h})-\tilde{h}/2]/[S_0/(2n\pi \tilde{h})+\tilde{h}/2]\}$). Since the local Weber number of the bubble is relatively low due to the small bubble size, the bubbles maintain their spherical{\color{black}-cap} shape although the bubbles could deform and the flows inside and outside the bubbles are both solved. {\color{black}The bubble shape is closer to spherical with larger $\tilde{h}$ and smaller $S_0/n$.}

To ensure that the contact lines are pinned on the plate, we set the value of $C$ on the hot bottom plate equal to the initial value as the boundary condition of the phase field. The other boundary conditions are no-slip velocities on the top and bottom plates, fixed temperature $\theta_{cold}=0$ (top) and $\theta_{hot}=1$ (bottom), and periodic conditions in the horizontal directions. Stretched grids with $288^3$ gridpoints are used for the velocity and temperature fields, and uniform grids with $576^3$ gridpoints for the phase field \citep{liu2021jcp}{\color{black}, which corresponds to at least $12$ gridpoints used for the bubble height}. The mesh is sufficiently fine and is comparable to corresponding single-phase studies \citep{ste10,poe13}. 

\begin{table}
\begin{center}
\begin{tabular}{ m{5.2cm} m{0.8cm}<{\centering} m{1.5cm}<{\centering} m{0.7cm}<{\centering} m{2cm}<{\centering} m{2cm}<{\centering} }
Material and geometrical parameters&&&($40^\circ C$)&Water&Air\\
\hline
\end{tabular}
\begin{tabular}{ m{4.2cm} m{1.8cm}<{\centering} m{1.5cm}<{\centering} m{0.7cm}<{\centering} m{2cm}<{\centering} m{2cm}<{\centering} }
Density&$\rho\,\,$&$kg/m^3$&&992.2&1.127\\[3pt]
Kinematic viscosity&$\nu\,\,$ &$m^2/s$&&$0.6591\times10^{-6}$& $16.92\times10^{-6}$\\[3pt]
Thermal conductivity &$k\,\,$& $W/(m\,K)$&&0.6286&0.02735\\[3pt]
Thermal diffusivity &$\kappa=k/(c_p\rho)\,\,$& $m^2/s$&&$1.52\times10^{-7}$& $241.0\times10^{-7}$\\[3pt]
Specific heat capacity &$c_p\,\,$ &$J/(kg\,K)$&&4180&1007\\[3pt]
Thermal expansion coefficient&$\beta\,\,$&$K^{-1}$&&$3.84\times10^{-4}$&$32.1\times10^{-4}$\\[3pt]
\end{tabular}
\begin{tabular}{ m{4.2cm} m{1.8cm}<{\centering} m{1.5cm}<{\centering} m{0.7cm}<{\centering} m{4.2cm}<{\centering}  } 
Surface tension&$\sigma\,\,$&$N/m$&&0.0696\\[3pt]
Gravity acceleration&$g\,\,$ &$m/s^2$&&9.8\\[3pt]
Temperature difference&$\Delta\,\,$ &$K$&&$0.1-10$\\[3pt]
Domain height &$H\,\,$&$m$&&$0.1-10$\\[3pt]
\hline
\end{tabular}
\begin{tabular}{ m{5cm} m{3cm} m{0.2cm}<{\centering}m{2.2cm}<{\centering} m{2cm}<{\centering}  } 
Dimensionless parameter&&&Realistic value&Present value\\
\hline
Rayleigh number&$\Ra=\beta_l g H^3 \Delta/(\nu_l \kappa_l)$&&$10^6-10^{14}$&$10^7-10^{8}$\\[3pt]
Prandtl number&$\Prn=\nu_l/\kappa_l$ &&4.35& 4.38\\[3pt]
Weber number &$\We=\rho_l U^2 H/\sigma$&&$10^{-1}-10^{5}$&100\\[3pt]
Froude number &$\Fr=U^2/(gH)$ &&$10^{-4}-10^{-2}$&1\\[3pt]
Density ratio&$\Lambda_\rho=\rho_g/\rho_l$&&0.001&0.01\\[3pt]
Kinematic viscosity ratio&$\Lambda_\nu=\nu_g/\nu_l$&& 25.7&25.7 \\[3pt]
Thermal conductivity ratio&$\Lambda_k=k_g/k_l$&&0.0435&0.042\\[3pt]
Thermal diffusivity ratio&$\Lambda_\kappa=\kappa_{g}/\kappa_{l}$ &&158&158\\[3pt]
Thermal expansion coefficient ratio&$\Lambda_\beta=\beta_g/\beta_l$&&8.36&8.36\\[3pt]
\hline
\end{tabular}
\end{center}
\caption{\label{table-para} The upper part of the table shows the material and geometrical parameters controlling the flow and their typical values. The lower part of the table shows all resulting dimensionless parameters. Here $U=\sqrt{\beta_l g H \Delta}$ is the free-fall velocity. The index $g$ stands for gas, and $l$ for liquid. 
}
\end{table}

\section{Flow features and heat transfer}
\label{sec1}

\begin{figure}
\centering
\includegraphics[width=\linewidth]{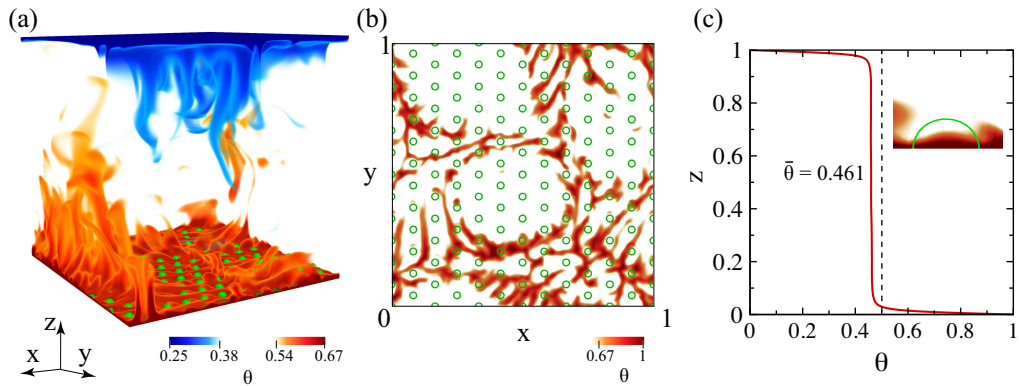}
\caption{\label{fig1} Rayleigh-B\'enard convection with bubbles on the hot plate for $\Ra = 10^8$, $\Prn=4.38$, $S_0=0.18$, $\tilde{h}=0.02$ and $n=144$. (a) Volume rendering of a snapshot of the thermal structures, (b) horizontal slice at the height of the lower (hot) boundary layer thickness, and (c) mean {\color{black}(over time and space)} temperature profile with an inset showing thermal structures inside one bubble. The temperature field is color-coded and the interface between liquid and gas is marked in green. As seen from the mean temperature profile in (c), due to the bubbles the mean centre temperature $\bar{\theta}=0.461$ is lower than the average temperature $1/2$ of the top and bottom plates (dashed black line).
}
\end{figure}


A typical thermal structure in RB convection with bubbles attached to the hot plate is shown in figure \ref{fig1}. Although there is continuous emission of thermal plumes, the bubbles almost maintain the shape of spherical cap due to the pinned contact lines and sufficiently strong surface tension. This is indeed the relevant situation during most of the time for the bubbles in electrolysis or catalysis. In figure \ref{fig1}(a) and (b), we observe the plumes rising up from the gaps between the bubbles. Inside the bubbles (see the inset in figure \ref{fig1}c), pure thermal conduction takes place since the local Rayleigh number inside the bubble $\Ra_g\approx\beta_g g h^3(\Delta/2)/(\nu_g\kappa_g)$ is small enough to remain under the onset of convection $\Ra_c\approx1708$ (namely $\Ra_g<100$) for all cases. Furthermore, considering that the thermal conductivity of gas is much lower than that of liquid (i.e. for their ratio $\Lambda_k=0.042\ll1$), the heat transfer through the gas phase is negligible. Therefore the overall heat-conducting ability of the fluid near the hot bottom plate is lower than that near the cold top plate. Consequently, since the total heat flux is the same across each horizontal plane, the temperature drop across the hot bottom BL becomes larger than that across the cold top BL, leading to the mean centre temperature smaller than $0.5$ (e.g. $0.461$ in figure \ref{fig1}c), closer to the temperature of the cold top plate to that of the hot bottom plate.

\begin{figure}
\centering
\includegraphics[width=\linewidth]{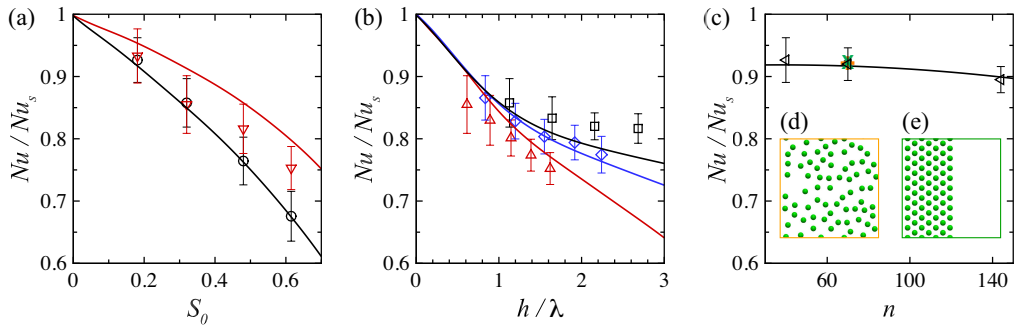}
\caption{\label{fig2}   Nusselt number normalized by that in the single-phase system. (a) $\Nun/\Nun_s$ as function of the relative bubble-covered area $S_0$. Here the (relative) bubble height $\tilde{h}=0.02$ and the bubble number $n=40$. (b)$\Nun/\Nun_s$ as function of $h/\lambda$ for $S_0=0.32$ and $n=40$, with $\lambda$ being the thermal BL thickness ({\color{black}values taken from the hot bottom BL}). (c)$\Nun/\Nun_s$ as function of $n$ for $S_0=0.18$ and $\tilde{h}=0.02$. Symbols denote the numerical results and lines the predictions of eqs. (\ref{lambda_hot}) and (\ref{equivalent})$-$(\ref{gl}), where colors are for $\Ra=10^7$ (red), $\Ra=3.2\times10^7$ (blue), and $\Ra=10^8$ (black). 
The orange symbol $+$ and the green symbol $\times$ in (c) denote the cases with random bubbles distribution as shown in (d) and the half-covered distribution as shown in (e). All the other symbols in (a)$-$(c) are for the uniform distribution. {\color{black}All error bars and deviations between simulations and predictions are within $5\%$.}}
\end{figure}

In figure \ref{fig2}, we plot the heat transfer $\Nun$ (normalized by $\Nun_s$ of single-phase system) as function of the geometrical parameters. As shown in figure \ref{fig2}(a), unsurprisingly, $\Nun/\Nun_s$ decreases with increasing the relative bubble-covered area $S_0$ due to the decreasing conducting area. We also observe that $\Nun/\Nun_s$ decreases with increasing bubble height $h$ (normalized by the thermal BL thickness $\lambda$) in figure \ref{fig2}(b), since for larger bubbles there is less liquid (the major conducting fluid) in the thermal BL. {\color{black}With increasing $h/\lambda$, the simulation results gradually deviate from the predictions, since the larger $h/\lambda$, the more parts of the bubbles enter the bulk, which gradually deviates from our assumption that bubbles only affect the thermal BL structure.}

In contrast to the significant effects of $S_0$ and $\tilde{h}$ on the heat transfer, $\Nun/\Nun_s$ is insensitive to the bubble number $n$ (see figure \ref{fig2}c), since $n$ only contributes to the second term in the bubble volume ($=S_0\tilde{h}/2+n\pi \tilde{h}^3/2$), which is of order $O(\tilde{h}^3)$ and thus negligible compared to the first term ($O(\tilde{h})$). We further show the effects of the spatial bubble distribution, including uniform, random, and the half bubble-covered distributions in figure \ref{fig2}(c). 
With all three types of distribution, the values of $\Nun/\Nun_s$ are almost the same. This indicates that the overall heat flux is insensitive to the spatial bubble distribution, at least in our model system. {\color{black}This differs from the previous studies \citep{liu2008prl,jiang2018prl}, where the large-scale flow is changed by the solid elements on the plate. The rigid solid surfaces imply no-slip velocity boundary conditions, which affect the flow structure much more than the boundary conditions of continuous velocity and shear stress on the deformable bubble interfaces.}

\begin{figure}
\centering
\includegraphics[width=\linewidth]{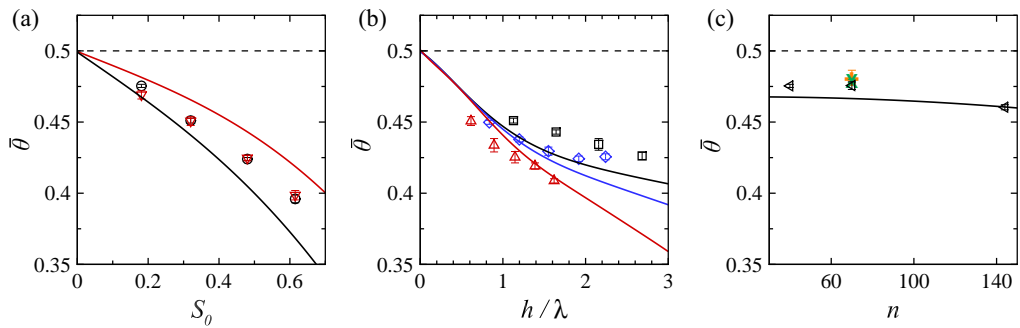}
\caption{\label{t-sh}   (a) Mean centre temperature $\bar{\theta}$ as function of $S_0$. (b) $\bar{\theta}$ as function of $h/\lambda$. (c) $\bar{\theta}$ as function of $n$. All the cases are the same as in figure \ref{fig2}. The dashed line denote the average temperature $1/2$ of the top and bottom plates. {\color{black}All error bars and deviations between simulations and predictions are within $5\%$.}}
\end{figure}

Analogous trends are also found in the relationship between the mean centre temperature $\bar{\theta}$ and the geometrical parameters, as shown in figure \ref{t-sh}. In all cases, the temperature profile is asymmetric due to the bubbles, namely $\bar{\theta}$ is lower than the average temperature $1/2$ between top and bottom plates, as explained above. {\color{black}The temperature profile can be quantitatively described by $\bar{\theta}$ and the top and bottom thermal BL thicknesses, the values of which are calculated in Section \ref{sec2}.}

Such an asymmetric temperature profile has already been studied in the context of single-phase RB convection under non-Oberbeck-Boussinesq (NOB) conditions \citep{ahl06}, where in water the viscosity and thermal diffusivity are temperature dependent and smaller at the hot bottom plate than at the cold top plate. Also this leads to an asymmetric temperature profile. \cite{ahl06} employed an extended Prandtl-Blasius BL theory for the NOB conditions in the two BLs, and coupled them by imposing heat flux conservation. With this they could calculate the centre temperature and the thermal BL thicknesses, which well agree with the experimental results. Note that that analytical calculation must be adapted to be applicable to the situation here, since here the BL flow is not parallel to the plates due to bubbles. In addition, whereas in the NOB case one has no-slip and no-penetration boundary conditions throughout, here on the spherical-cap shaped bubble interface we have the emerging condition of the velocity and shear stresses continuity. However, as in \cite{ahl06}, we use the concept of heat flux conservation to define the thermal BL thickness (see section \ref{sec-bl}), and then propose the idea of an equivalent single-phase system to mimic the system with attached bubbles (see section \ref{sec2}).

\section{Thermal BL thicknesses}
\label{sec-bl}

\begin{figure}
\centering
\includegraphics[width=0.9\linewidth]{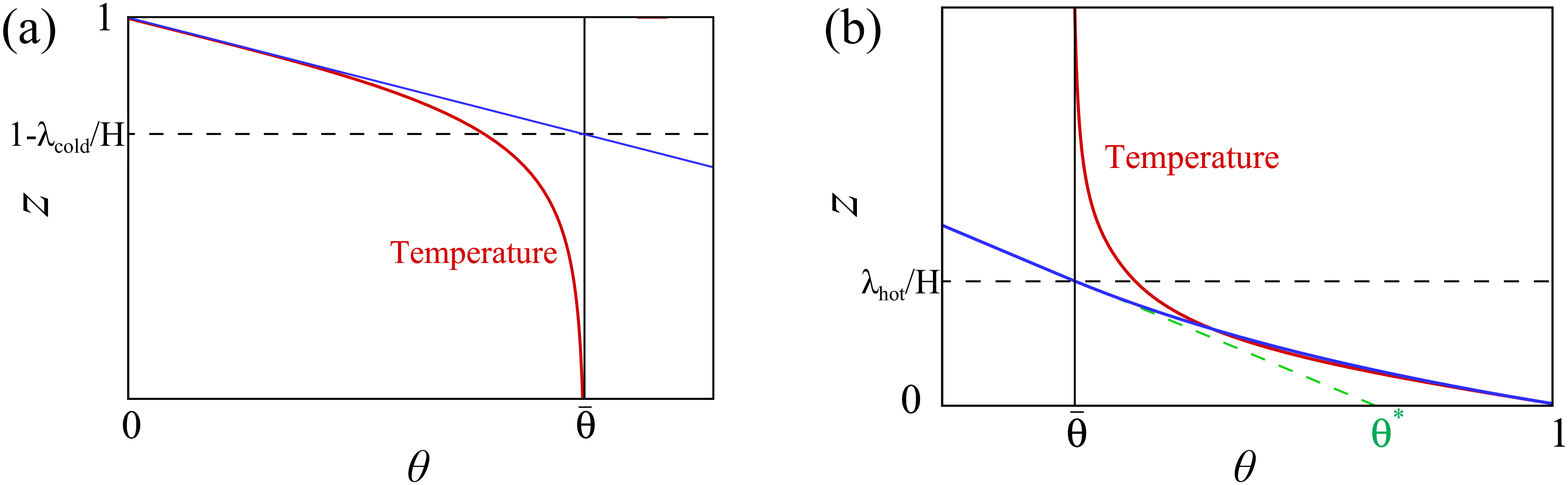}
\caption{\label{fig4} Sketch of the temperature profile near the (a) cold top and (b) hot bottom plate. The red lines denote the temperature profiles, the black lines the mean centre temperature $\bar{\theta}$, the blue lines the solution for pure thermal conduction in the BL, i.e. eq. (\ref{lambda_hot}), and the green dashed line the tangent of (\ref{lambda_hot}) at the intersection. $\theta^*$ is the intercept of the tangent. We define the dimensional BL thickness near the hot and cold plates as $\lambda_{hot}$ and $\lambda_{cold}$, respectively, i.e., the distance from the black dashed lines to the corresponding plate.
}
\end{figure}

A sketch of the temperature profiles near the plates is displayed in figure \ref{fig4}, where we also show how the thermal BL thicknesses are defined. Near the cold top plate (see figure \ref{fig4}a), the thermal BL thickness $\lambda_{cold}$ is defined through the usual convenient definition via the slope of the temperature profile at the plate from assuming a pure conductive thermal BL and a well-mixed bulk \citep{ahl06}. As $\lambda_{cold}$ we take that distance from the plate, where the tangent to the temperature profile at the plate reaches the mean centre temperature $\bar{\theta}$. 
However, this definition cannot be directly applied near the hot bottom plate, due to the bubbles. As the bubbles reduce the area occupied by the liquid, the conducting area varies with $z$. We therefore correct the solution for pure thermal conduction in the hot bottom BL based on heat flux $Q$ conservation,
\begin{equation}
    Q=k_l\frac{\partial \theta}{\partial z}[1-S(z,S_0,\tilde{h},n)]=k_l\frac{\bar{\theta}H}{\lambda_{cold}}.
    \label{lambda_hot}
\end{equation}
Here, the term on the right is the heat flux on the cold top plate. The gas-covered area $S(z,S_0,\tilde{h},n)=max[(S_0+n\pi z \tilde{h})(1-z/\tilde{h}),0]$ is the insulating area, which is defined with the assumption of spherical-cap shaped bubbles. 

\begin{figure}
\centering
\includegraphics[width=\linewidth]{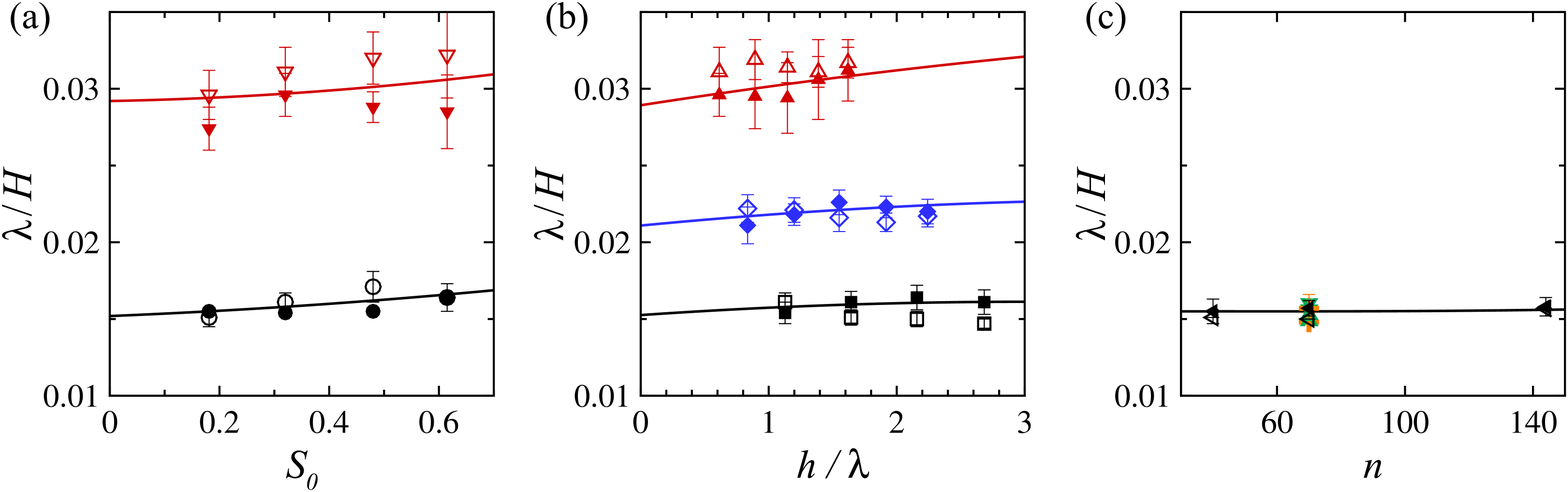}
\caption{\label{lambda-sh}   (a) Thermal BL thicknesses $\lambda_{hot}$ (empty symbols) and $\lambda_{cold}$ (filled symbols) of the hot bottom and cold top BLs, respectively, as functions (a) of $S_0$, (b) of $h/\lambda$, and (c) of $n$. All the cases are the same as in figure \ref{fig2}. {\color{black}All error bars and deviations between simulations and predictions are within $5\%$.}}
\end{figure}

The thermal BL thickness near the hot bottom plate $\lambda_{hot}$ equals the distance from the plate to the intersection between the lines of (\ref{lambda_hot}) and of $\theta=\bar{\theta}$, as shown in figure \ref{fig4}(b). With the definitions above, both values of $\lambda_{hot}$ and $\lambda_{cold}$ are close to each other in all cases (within $10\%$), as shown in figure \ref{lambda-sh}. {\color{black}Since $\lambda_{hot}$ and $\lambda_{cold}$ are almost the same and $\bar{\theta}<(\theta_{hot}+\theta_{cold})/2$, the temperature variations across the hot bottom BL and the cold top BL are different, which is reflected in that the temperature profile is asymmetric.}


\section{Predictions for $\Nun$ and the centre temperature using equivalent single-phase RB system}
\label{sec2}

\begin{figure}
\centering
\includegraphics[width=\linewidth]{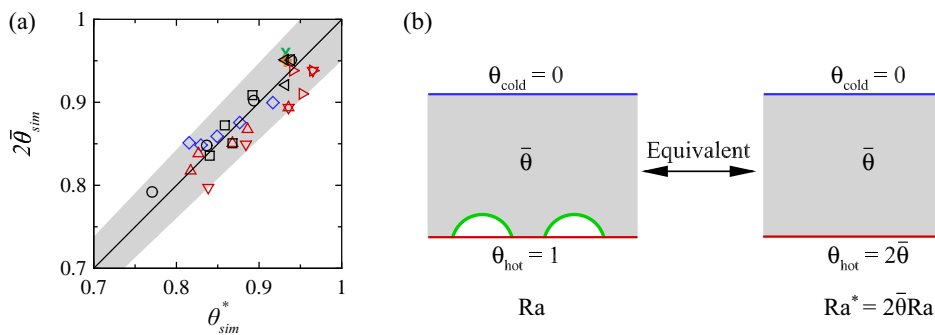}
\caption{\label{fig5} (a) Comparison {\color{black}between the mean centre temperature $\bar{\theta}_{sim}$ and the effective temperature $\theta^*_{sim}$, both taken from simulations.} The symbols are the same as in figure \ref{fig2}. The shadow is the zone within $\pm 5\%$. (b) Sketch of the bubble-system under consideration and the equivalent single-phase RB system. {\color{black} $\bar{\theta}_{sim}$ and $\theta^*_{sim}$ are plotted this way to support the existence of the equivalent single-phase system.}}
\end{figure}

To calculate the heat transfer $\Nun$ and the mean centre temperature $\bar{\theta}$, we propose the idea of using an equivalent single-phase setup to mimic the system with attached bubbles. To find the equivalent flow, we first obtain the effective temperature at the hot bottom plate. This is done by plotting the tangent of (\ref{lambda_hot}) at the position of $\lambda_{hot}$, and the obtaining $\theta$-intercept of the tangent as effective temperature $\theta^*$ (see figure \ref{fig4}b). 

Next, we compare the effective temperature $\theta^*$ and the mean centre temperature $\bar{\theta}$ in figure \ref{fig5}(a), which shows that $\theta^*$ approximately equals to $2\bar{\theta}$ (within $\pm5\%$). Thus we assume a nearly equivalent single-phase RB counterpart with $\theta_{hot}=2\bar{\theta}$ and $\theta_{cold}=0$, such that $\bar{\theta}$ is the mean value of temperatures on the two plates, as shown in figure \ref{fig5}(b). Again, we note that $\lambda_{hot}$ is close to $\lambda_{cold}$, based on our definition of the thermal BL thickness in Section \ref{sec1}. Thus, we have the following relations:
\begin{equation}
    \theta^*=2\bar{\theta},\quad\lambda_{cold}=\lambda_{hot}={\lambda_s}.
    \label{equivalent}
\end{equation}
Here $\lambda_s$ is the thermal BL thickness in the single-phase system, which can be calculated as
\begin{equation}
    \lambda_s=\frac{\bar{\theta}H}{\Nun_e},
\end{equation}
where $\Nun_e$ is the heat transfer estimated from the GL theory \citep{gro00,gro01} for the equivalent single-phase system with $\Ra^*=2\bar{\theta}\Ra$, $\theta_{hot}=2\bar{\theta}$ and $\theta_{cold}=0$. In the two-phase system and the equivalent single-phase system, the dimensional heat transfer $Q$ ($=\Nun\,k(\theta_{hot}-\theta_{cold})/H$) should be the same, which yields
\begin{equation}
    \Nun(\Ra,\Prn,S_0,\tilde{h},n)=2\bar{\theta} \Nun_e(\Ra^*,\Prn),
    \label{gl}
\end{equation}
where $\Nun(\Ra,\Prn,S_0,\tilde{h},n)=(\bar{\theta}H)/\lambda_{cold}$ is the heat transfer for the two-phase system.

Combining the relations (\ref{equivalent})$-$(\ref{gl}) for the equivalent single-phase RB system with the relationship (\ref{lambda_hot}) between $\bar{\theta}$ and $\lambda_{cold}$ in section \ref{sec-bl}, we can now calculate the heat transfer in the system with attached bubbles.



We emphasize that with this approach, using the equations above without introducing any free parameter, for given $\Ra$, $\Prn$ and bubble geometries, we can now calculate $\Nun$, $\bar{\theta}$, $\lambda_{hot}$ and $\lambda_{cold}$. The good agreements between the simulations and the predictions for $\Nun$, $\bar{\theta}$, $\lambda_{hot}$, and $\lambda_{cold}$ are shown in figures \ref{fig2}, \ref{t-sh}, and \ref{lambda-sh}, respectively, where all deviations between simulations and predictions are within $\pm 5\%$. 



\begin{table}
\begin{center}
\begin{tabular}{ m{2cm} m{2cm}<{\centering} m{2cm}<{\centering} m{2cm}<{\centering} m{2cm}<{\centering} m{2cm}<{\centering} }
&$\tilde{h}$&$\Nun/\Nun_s$&$\bar{\theta}$&$\lambda_{bot}/H$&$\lambda_{top}/H$\\
\hline
Simulations&0.03&0.900$\pm$0.030&0.462$\pm$0.005&0.0303$\pm$0.0023&0.0294$\pm$0.0008\\[3pt]
Predictions&0.03&0.880&0.452&0.0340&0.0340\\[1pt]
\hline
Simulations&0.05&0.821$\pm$0.022&0.438$\pm$0.003&0.0315$\pm$0.0180&0.0299$\pm$0.0006\\[3pt]
Predictions&0.05&0.801&0.420&0.0347&0.0347\\[3pt]
\end{tabular}
\end{center}
\caption{\label{table-pr} Comparisons between simulations and predictions at large $\Prn=400$. The other parameters are $\Ra=10^7$, $S_0=0.32$, and $n=40$.}
\end{table}


We further check whether our approach is also applicable for large $\Prn$, which, as explained above, has relevance to transport phenomena in water electrolysis and catalysis. Water electrolysis and catalysis can both lead to natural convection driven by buoyancy, which originates from the density difference of the solute with different concentrations of the electrolysis or catalysis product. Here, the mass transfer is also characterized by $\Nun$, i.e. the mass transfer normalized by that with pure diffusion (in this context normally called Sherwood number $Sh$). The control parameters are the Grashof number (dimensionless strength of the solute driving) $Gr$ and the Schmidt number (the ratio of viscous diffusion and mass diffusion rates) $Sc$, corresponding to $\Ra/\Prn$ and $\Prn$ in RB convection, respectively. The value of $Sc$ in water electrolysis is always large, e.g., $Sc=400$ \citep{sepahi2021}.

We performed two simulations (for two different bubble heights) at $\Prn=400$ and $\Ra=10^7$ (tabulated in Table \ref{table-pr}) with a sufficiently fine mesh as explained in section \ref{meth}. Note the much higher computational costs at this large $\Prn$, due to the required long time ($\sim \Prn^{1/2}$) for the system to enter the statistical steady state. The bubble geometries are characterized by $\tilde{h}=0.03$ and $0.05$, $S_0=0.32$, and $n=40$. The good agreements between our parameter free predictions and the results from the simulations are shown in Table \ref{table-pr}. This supports that our predictions can be directly applied to water electrolysis and catalysis.

\section{Conclusions and outlook}\label{conc}

Turbulent RB convection with gas bubbles attached to the hot plate is numerically investigated for  $10^7\le \Ra\le 10^8$ and $\Prn=4.38$ and $400$. The bubble geometrical parameters are the relative bubble-covered area $S_0$, the relative bubble height $\tilde{h}$, the bubble number $n$, and the spatial bubble distribution. Due to the much lower thermal conductivity of gas as compared to liquid, the temperature profile is asymmetric and the heat transfer efficiency of the system is reduced. More specifically, $\Nun$ significantly decreases with increasing $S_0$ and $\tilde{h}$, but is almost unaffected by $n$ and the types of bubble distribution. 

To predict the heat transfer and the mean centre temperature of the system, we have proposed the idea of using an equivalent single-phase system to mimic the system with attached bubbles. By applying the GL theory for the equivalent system and imposing heat flux conservation in the two thermal BLs, we can predict the heat transfer, the top and bottom thermal BL thicknesses, and the mean centre temperature, without introducing any free parameter. The predictions well agree with the results from the simulations. {\color{black}Briefly, in Section \ref{sec-bl} we got one relationship between $\bar{\theta}$ and $\lambda$ in eq. (\ref{lambda_hot}), and in Section \ref{sec2} we got another relationship between $\bar{\theta}$ and $\lambda$ in eqs.(\ref{equivalent}) and (\ref{gl}). Then, for only given $\Ra$, $\Prn$ and bubble geometries, we can well predict the heat transfer and temperature profile in the system with bubbles attached to the bottom plate.}

The results of this study can be used not only for the heat transfer in RB convection with bubbles attached to the plates, but also for the mass transfer in electrolysis or catalysis. Our predictions can help to obtain estimates for relevant applications and e.g. optimize the heat or mass transfer and flow features in systems in which bubbles are forming on the plate(s). It would also be interesting to extend our basic idea to other wall-bounded turbulent systems with various plate properties, such as plates with inhomogeneous properties (e.g. wettability or conducting ability).


\section*{Acknowledgments}
We acknowledge PRACE for awarding us access to MareNostrum in Spain at the Barcelona Computing Center (BSC) under the project 
$2021250115$ and the Netherlands Center for Multiscale Catalytic Energy Conversion (MCEC). K. L. Chong acknowledges Shanghai Science and Technology Program under project no. 19JC1412802.

\section*{Declaration of interests}
The authors report no conflict of interest.


\bibliographystyle{jfm}
\bibliography{liuhr,literatur,2ph_literatur}

\end{document}